\documentclass[10pt]{iopart}

\usepackage{graphicx}
\usepackage{ulem} 
\usepackage{dcolumn}
\usepackage{bm}
\usepackage[colorlinks=true, citecolor=blue]{hyperref}
\usepackage{array}
\usepackage{float} 
\usepackage{silence}
\usepackage{cite}
\usepackage{booktabs}
\usepackage{epstopdf}
\usepackage[utf8]{inputenc}
\usepackage{ragged2e}
\usepackage{hyperref}
\usepackage[mathlines]{lineno}

\usepackage{iopams}  

\begin{document}

\title[]{Effect of LaNiO$_{3}$ on the impedance and dielectric properties of CoFe$_{2}$O$_{4}$: a high temperature study}

\author{Ananya Patra$^1$$^,$$^*$ and V Prasad$^1$}
\address{$^1$ Department of Physics, Indian Institute of Science, Bangalore 560012, Karnataka, India}
\ead{$^*$ananyap@iisc.ac.in}
\vspace{10pt}
\begin{indented}
\item[]March 2019
\end{indented}

\begin{abstract}

 We report the high temperature impedance and dielectric measurement of the nanocomposites comprising of CoFe$_{2}$O$_{4}$ (CFO) and LaNiO$_{3}$ (LNO) by changing the content of LaNiO$_{3}$ (0, 5, 10 and 15 \%). The aim of the present study is to relate the structural defects expected at the interfaces of the composites with the result of impedance and dielectrics. The result separately shows the grain, grain boundary and electrode relaxation behaviour in impedance and modulus spectroscopy. As calculated using Arrhenius law it indicates a reduced activation energy of the grain boundary in the presence of LNO compared to pure CFO. This assists the charge carriers for short range hopping across the boundary as reflected from ac conductivity. The dielectric constant at high frequency side manifests the interband transition in both pure CFO and the composite materials. The frequency of the transition can be tuned by LNO content. Thus this detail study will help to understand the change of relaxation phenomenon and dielectric behaviour in the presence of structural defects and disorder.  
\end{abstract}
%
\noindent{\it Keywords\/}: nanocomposites, impedance, relaxation, dielectric, interband transition, ac conductivity

%
%
%
\ioptwocol
\section{Introduction}
Composites are one of the most promising materials where the combination of two or more constituents shows remarkable mechanical and physical properties while retaining the phase of individual components. The advantage of tuning the properties by varying compositions have made the composites very popular materials in the field of multiferroics \cite{1,2}, superconductivity \cite{3,4} or to improve mechanical and thermal properties \cite{5} or magneto transport properties \cite{6,7}. From our previous studies, we have seen that the structural defects have huge impact on the transport properties of the composites containing LaNiO$_{3}$ (LNO) and CoFe$_{2}$O$_{4}$ (CFO) \cite{8}. Therefore, it will be interesting to correlate the electrical properties with the microstructure of the composite materials. For that purpose we have used complex impedance spectroscopy which is a powerful tool to determine various relaxation process present in grain, grain boundaries and electrode-interfaces of polycrystalline materials. CFO is one of the widely used spinel ferrites because of its properties as hard ferrimagentic material, high electrical resistance, good chemical and thermal stability and large magnetic anisotropy. 
There are previous reports on the impedance and dielectric studies of pure CFO in nanocrystalline form \cite{9} or in thick films \cite{10} 
and on the composites of CFO and other ferroelectric materials for example BiFeO$_{3}$ \cite{11,12} or BaTiO$_{3}$ \cite{13} in an attempt to produce multiferroic materials. The purpose of our present study is to thoroughly analyse the various relaxation processes in pure CFO in nanocrystalline form and how they change in the presence of a material of different conductivity and dielectric constant compared to CFO.

In  our previous report, we have seen that the defects inside the crystalline planes and at the interfaces between these two crystalline phases influence the charge transport \cite{8}. Therefore we can expect these defects will help to trap more space charges and will effect the impedance and dielectric properties. The composites are prepared by the insulating matrix of CFO with conductive filler LNO by changing LNO content (0, 5, 10 and 15 \%). As expected, the result of impedance and dielectric properties reflect the effect of space charges at both grain and grain boundary. The dielectric constant of pure CFO shows an unique nature showing negative value at high frequency side ($\sim$ 90 MHz at 35 $^{\circ}$C) which corresponds to interband transition. At the best of our knowledge this is the first report on interband transition of CFO at such low frequency which probably results from the nanocrystalline size of CFO. The ac conductivity reflects the direct consequence of the reduced activation energy which is manifested in the composite with 15 \% LNO. Therefore, we believe this elaborate study will help us to understand the dielectric properties and ac conductivity in the presence of disorder and defects.  


\section{\label{sec:level2 }Experimental Details}

The nanoparticles (nps) of LaNiO$_{3}$ (LNO) and CoFe$ _{2} $O$ _{4} $ (CFO) have been synthesized by citric acid assisted sol-gel method \cite{8}. 
The two powders of LNO and CFO are mixed and ground to get a homogeneous mixture and heated at 650 $ ^{\circ} $C. To measure dielectric properties, the heat treated powder is pressed into pellet and sintered at 700 $ ^{\circ} $C.

The Samples are characterized by X-ray diffraction (XRD) and scanning electron microscopy (SEM). XRD is performed using Rigaku Smartlab diffractometer. SEM image is collected with Ultra 55 FE-SEM Karl Zeiss. The magnetic property is measured using MPMS SQUID magnetometer equipped with magnetic field upto 7 T. The impedance spectroscopy is measured in the frequency range 100 Hz to 110 MHz in the temperature range 35--400 $^{\circ}$C using Agilent 4294A precision impedance analyzer.

\section{\label{sec:level3} Results and discussion}

\subsection{\label{sec:level4} X-ray diffraction}
\begin{figure}[htbp]
	\centering
	\includegraphics[width=1\linewidth]{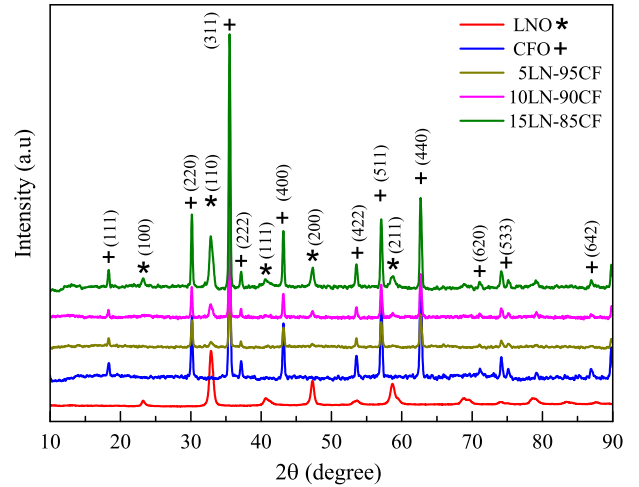}
	\small{\caption{XRD pattern of the composites xLNO + (1-x)CFO where x = 0, 0.5, 0.10, 0.15, 0.20 and 1 \label{fig.1}}}
\end{figure} 
Figure~\ref{fig.1} shows the X-ray diffraction pattern of the series [xLNO +(1-x)CFO (x = 0, 0.05, 0.10, 0.15, 0.20 and 1)] which confirms the phase formation of the two materials LNO and CFO and the phase purity of their composites. The independent presence of the peaks of LNO and CFO (indicated by * and + respectively) reveals that both materials co-exist in the composites. Although the CFO peaks are more prominent than LNO because of higher percentage of CFO. The diffraction peaks of CFO correspond to cubic spinel structure with space group Fd$\bar{3}$m (JCPDS No. 22-1086). Since there is no peak splitting of LNO in pure form and also in the composites, the peaks of LNO are indexed by the cubic perovskite structure with space group Pm$\bar{3}$m (JCPDS No. 33-0710). We have calculated the crystallite size of LNO and CFO using Scherrer equation. The variation of average crystallite size of LNO in pure form and in composites is from 14 nm to 19 nm and that of CFO is from 34 nm to 38 nm. The lattice parameters ($a$) of pure LNO and CFO as obtained from the relation: $a$ = $d_{hkl} \sqrt{h^{2}+k^{2}+l^{2}}$ are 3.846 \AA and 8.371 $^{o}$A respectively, where $d_{hkl}$ is the lattice spacing and (h,k,l) are the Miller indices. The lattice parameter does not change much in the composites.
\begin{figure*}[htbp]
	\centering
	\includegraphics[width=0.8\linewidth]{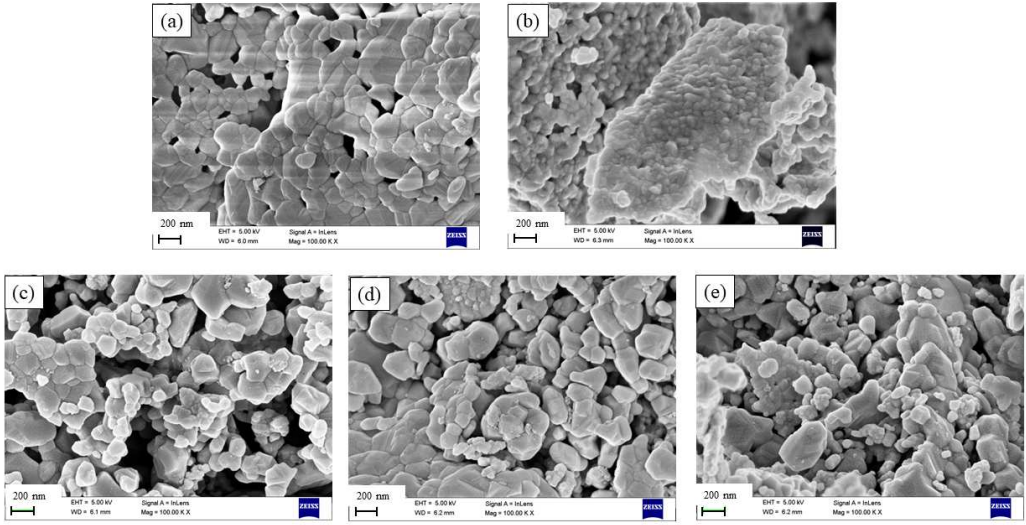}
	\small{\caption{SEM images of (a) pure CFO, (b) pure LNO and the composites: (c) 5LN-95CF, (d) 10LN-90CF and (e) 15LN-85CF. \label{fig.2}}}
\end{figure*} 
\subsection{\label{sec:level5} SEM}

The morphologies of pure and composite materials [xLNO +(1-x)CFO (x = 0, 0.05, 0.10, 0.15, 0.20 and 1)] are investigated by scanning electron microscopy as presented in Figure~\ref{fig.2}. 
The samples are formed of grains separated by grain boundaries which is also evident from the result of impedance spectroscopy as discussed in Section~\ref{sec:level7}. As it depicts from the figure, the grain size of pure CFO is little bit higher than pure LNO. Hence the small grains in the composites belong to LNO and the large grains are formed by CFO.

\subsection{\label{sec:level6} Magnetic property}

\begin{figure}[htbp]
	\centering
	\includegraphics[width=1\linewidth]{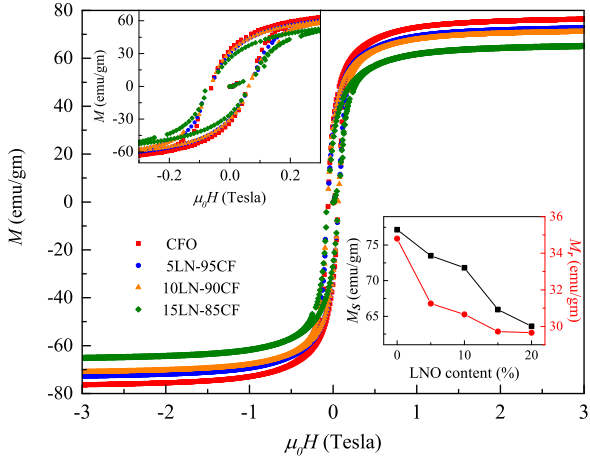}
	\small{\caption{ Magnetization as a function of magentic field at room tempeature (300 K) of pure CFO and the composites with LNO content 5, 10 and 15 \%. Top inset shows the enlarge view at low magentic field and the bottom inset depicts the variation of satuaration and remenant magnetization with LNO content.\label{fig.3}}}	
\end{figure} 

To investigate the magnetic properties, we have obtained the hysteresis loop (magnetization ($M$) vs. magnetic field ($\mu_{0}H$)) at room temperature for different composites [xLNO +(1-x)CFO (x = 0, 0.05, 0.10, 0.15, 0.20)] as shown in \Fref{fig.3}. The top inset represents the enlarge view at low magnetic field which shows there is not much change in the coercivity for different composites. However, the values of saturation ($M_{S}$) and remanent magnetization ($M_{r}$) are reduced consistently with decreasing CFO content in the composites as shown in the bottom inset of \Fref{fig.3}. The value of $M_{S}$ for pure CFO is close to that obtained for nano-sized CFO with particle size 48 nm \cite{14}. This particle size is consistent with the crystallite size determined from Scherrer equation (\Sref{sec:level4}).

\subsection{\label{sec:level7} Electrical analysis of pure CFO}

In polycrysttaline samples, both grain and grain boundary have significant contributions on the electrical properties. In impedance spectroscopy ($Z^{''}$ vs. $log(f)$), the low frequency effects are more prominent whereas the modulus spectroscopy ($M^{''}$ vs. $log(f)$) shows the high frequency effects more clearly \cite{15,16}. Therefore these two spectroscopy are complimentary to each other. The grain boundary effects dominates at low frequency and high frequency is mostly influenced by the bulk effect. Hence we have used both impedance and electric modulus spectroscopy to analyse the contributions from grain and grain boundary distinctly.

\subsubsection{\label{sec:level8} Impedance spectroscopy}


\begin{figure}[htbp]
	\centering
	\includegraphics[width=0.9\linewidth]{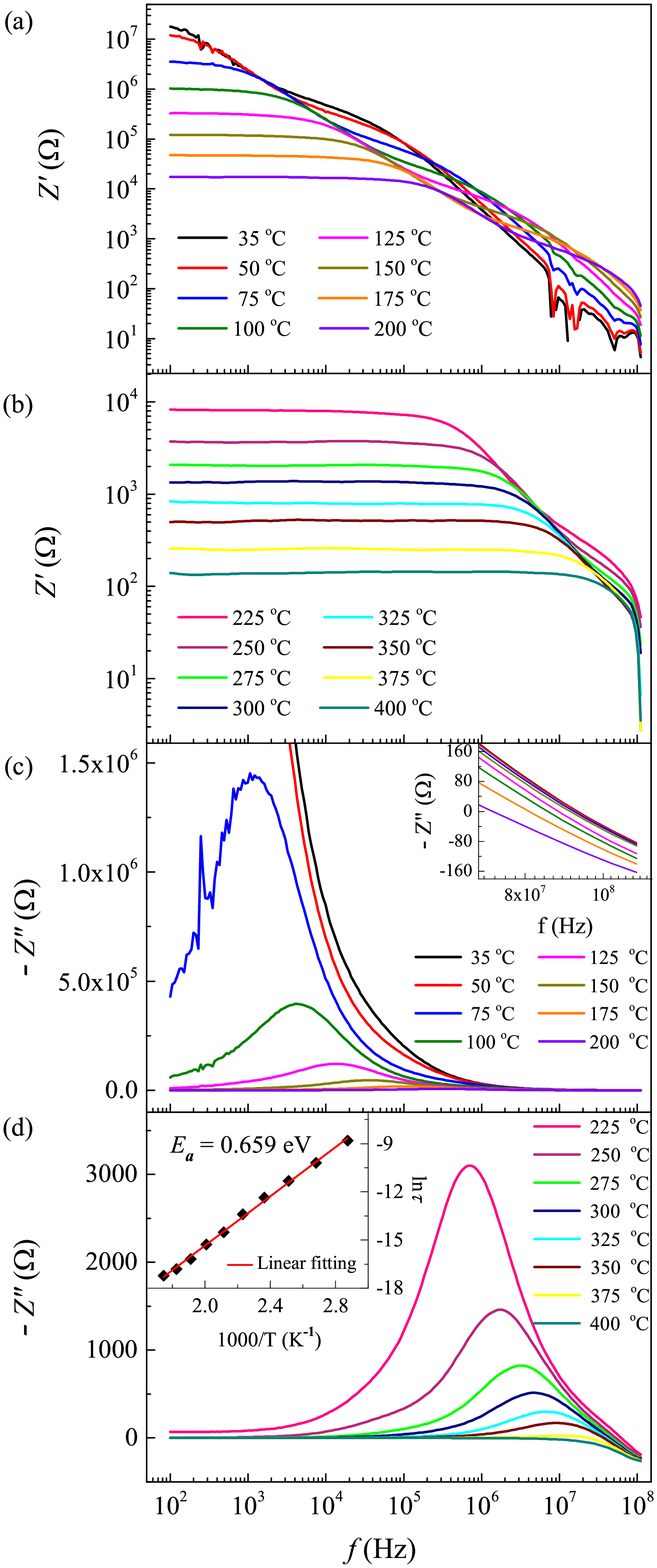}
	\small{\caption{Impedance analysis of pure CFO: real part of impedance in the temperature range (a) 35 -- 200 $^\circ$C, (b) 225 -- 400 $^\circ$C and imaginary part of impedance for (c) 35 -- 200 $^\circ$C, (d) 225 -- 400 $^\circ$C. Inser of (d) shows the relaxation time following Arrhenius law. \label{fig:Z}}}
\end{figure}

Fig.~\ref{fig:Z} (a) and (b) represents the real part ($Z'$) of impedance as a function of frequency at two different temperature regimes, $T \leq$ 200 $^{\circ}$C and 200 $^{\circ}$C $\leq T \leq$ 400 $^{\circ}$C respectively. $Z'$ starts to decrease with increasing frequency followed by a region at the high frequency side ($f >$ 10$^ {6}$ Hz) where all the graphs of different temperatures merge on each other. The decrease of $Z'$ with increasing frequency could be related to the space charge polarization \cite{17} which may be generated from carriers or Oxygen vacancies trapped in the bulk, at the grain boundaries in polycrystalline materials, at the interfaces between two materials or may be injected from electrode \cite{18}. 
At high frequency, the merging of $Z'$ at all temperature suggests 
the release of the space charges as a result of lowering the barrier properties in the ceramic materials
\cite{19,20,21}. The decrease of $Z'$ with increasing temperature indicates negative temperature coefficient of resistance and semiconducting like behaviour. The transport in ceramics mainly occur via the nearest neighbour hopping of charge carriers. With increasing temperature, more thermal energy is available for hopping across the potential barrier created by lattice defects or the neighbouring atoms or ions which leads to decrease in $Z'$ \cite{21,22}. 

The frequency variation of $Z''$ (Bode plot) as shown in Fig.~\ref{fig:Z} (c) (for $T \leq$ 200 $^{\circ}$C) and (d) ( for 200 $^{\circ}$C $\leq T \leq$ 400 $^{\circ}$C) reveals that $Z''$ is negative at low and intermediate frequency  which indicates capacitive nature of CFO and a small positive tail exists at the high frequency side ($\geq$ 90 MHz at 35 $^{\circ}$C) as shown in the inset of Fig.~\ref{fig:Z} (c). This positive part slowly grows larger as the temperature is increased and becomes more prominent above 200 $^{\circ}$C. 
This behaviour indicates a switching from capacitive to inductive reactance at high frequency region and an enhanced inductive effect with rising temperature. The capacitor may be formed when a thin layer of dielectric material is trapped between the comparatively conductive regions. This is called micro-capacitor \cite{23}. As the frequency is increased  due to the release of space charges and hence free electron conduction, the network of micro-capacitors breaks and forms the current loops equivalent to the inductor \cite{24,25}. 
A distinct peak is present at all temperatures which represents relaxation phenomena. The peak shifts to high frequency with reduced intensity when the temperature is increased from 35--400 $^{\circ}$C   
which implies thermally activated relaxation behaviour. The full width at half maxima of the peak is greater than the ideal Debye relaxation value (1.141 decade) and also the peak is asymmetric which indicate non-Debye type relaxation phenomenon. The activation energy associated with the relaxation behaviour is calculated using the Arrhenius law:

\begin{equation}
\tau = \tau_{0}exp(-E_{a}/k_{B}T),   
\label{eq.1}
\end{equation}

where $\tau$ = 1/2$\pi f_{P}$, $f_{P}$ is the frequency at which the peak occurs, $\tau_{0}$ is the characteristic relaxation time, $E_{a}$ is the activation energy 
and $k_{B}$ is the Boltzmann constant. The value of $E_{a}$ and $\tau_{0}$ as determined from the fitting in the inset of Fig.~\ref{fig:Z} are 0.659 eV and 4.86 $\times$ 10$^{-14}$ Sec, respectively. 

\begin{figure*}[htbp]
	\centering
	\includegraphics[width=0.8\linewidth]{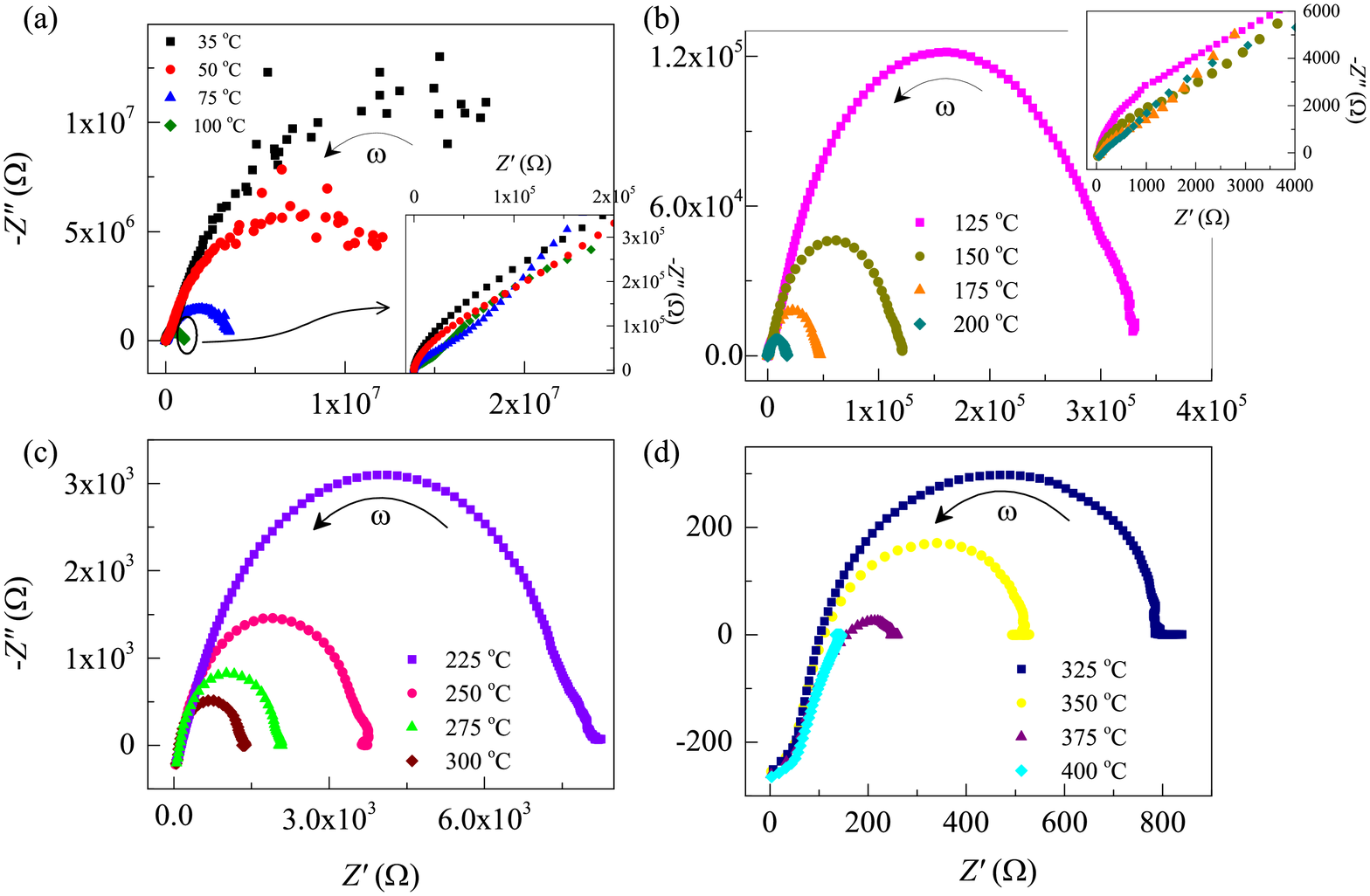}
	\small{\caption{Nyquoist plot of pure CFO in four different temperature ranges
			(a) 35 -- 100 $^\circ$C, (b) 125 -- 200 $^\circ$C, (c) 225 -- 300 $^\circ$C and (d) 325 -- 400 $^\circ$C   
			\label{fig:Z12}}}
\end{figure*}
To understand the nature of the relaxation associated with this peak we have shown Nyquist plot ($Z''$ vs. $Z'$) at different temperature regions (Figure~\ref{fig:Z12}). It shows a broad semicircular arc at low frequency which becomes full semicircle at $\sim$ 100 $^{\circ}$C. With a detail inspection a segment of a weak semicircle can be observed at high frequency as shown in the inset of Fig.~\ref{fig:Z12} (a) and (b). This small arc becomes weaker with increasing temperature and ultimately disappears at $\geq$ 200 $^{\circ}$C. Usually in dielectric materials, the electrical response of the grain boundary is related to larger resistance and capacitance than that of grain. This may be because in metal oxides, sintering the sample at high temperature can create Oxygen vacancies and during cooling down the sample at room temperature re-oxidation process may occur at the grain boundary which leads to insulating nature of the grain boundary whereas due to the presence of Oxygen deficient ions in grains, it can be conducting \cite{21,26}. Hence the low frequency relaxation corresponds to the grain boundary effect and the high frequency response is related to grain \cite{27,28}. 
Therefore the broad semicircle belongs to the relaxation of the space charges accumulated by the grain boundary and the small weak semicircular arc is associated with the space charge relaxation inside the grains. The impedance formalism mostly emphasizes the grain boundary conduction process, while grain effects would dominate in the electric modulus formalism. Therefore to extract the information about grain effect, we have considered modulus spectroscopy which is complimentary to impedance spectroscopy in the next section.

\subsubsection{\label{sec:level9} Modulus spectroscopy}

The electric modulus is also an complex quantity which can be expressed as below:

\begin{equation}
M^{*} = M'+ jM''= j\omega C_{0}Z^{*}
\label{eq.2}
\end{equation}  
and the imaginary ($M''$) part of electric modulus is,

\begin{equation}
M'' = \omega C_{0}Z'
\label{eq.3}
\end{equation}

Where $C_{0}$ is the geometric capacitance given by, $C_{0} = \frac{\varepsilon_{0}A}{t}$.
$A$ is area and $t$ is the thickness of the sample.

\begin{figure}[htbp]
	\centering
	\includegraphics[width=0.9\linewidth]{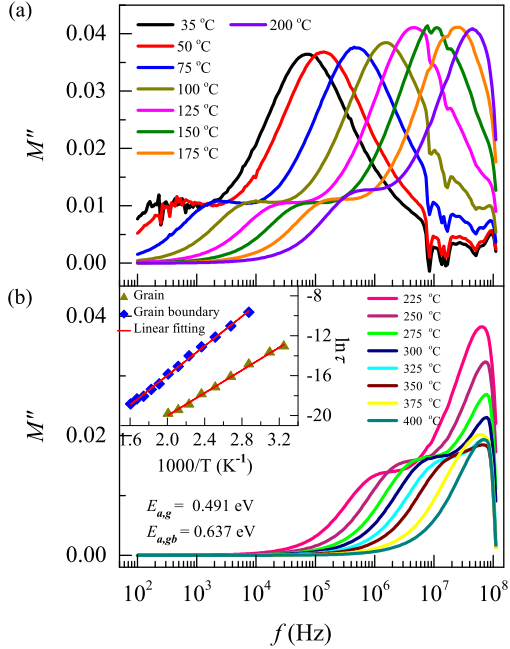}
	\small{\caption{Electric modulus of pure CFO: real part of modulus in the temperature range (a) 35 -- 200 $^\circ$C, (b) 225 -- 400 $^\circ$C and imaginary part of modulus for (a) 35 -- 200 $^\circ$C, (b) 225 -- 400 $^\circ$C.  Inset of (d) shows the linear relation between ln$\tau$ and 1/$T$ for grain and grain boundary. 
			\label{fig:M}}}
\end{figure}

Fig.~\ref{fig:M} (a) and (b) present the imaginary part of modulus ($M''$) with frequency. 
Two peaks are clearly visible in $M''$ at low and high frequency. Both peaks shift to high frequency side with rising temperature and 
above 350 $^{\circ}$C, the two peaks start to overlap and a only single peak is observed above that temperature.  This may be because at that high temperature the grain and grain boundary do not exist individually and the sample as a whole shows a single relaxation behaviour.
The activation energy is determined in similar way as described in last section (inset of Figure~\ref{fig:M} (d)) and the values for grain ($E_{a,g}$) and grain boundary ($E_{a,gb}$) are 0.491 eV and 0.637 eV, respectively. The value of $E_{a,gb}$ also matches with that calculated from impedance analysis which confirm the peak in $Z''$ is due to grain boundary effect. The characteristic relaxation time are 2.27 $\times$ 10 $^{-14}$ sec and 4.37 $\times$ 10$^{-14}$ sec for grain and grain boundary, respectively. 


To interpret the impedance data, an equivalent circuit is required that can represent the electrical properties of the dielectric materials. For polycrystalline materials, the equivalent circuit usually consists of series combination of two parallel resistance ($R$)-capacitance ($C$) elements. Those two parallel $R-C$ elements represent grain and grain boundary effects separately. The magnitude of the peak in impedance plot ($Z''$ vs. $f$) and in modulus plot ($M''$ vs. $f$) are proportional to $R$ and $\frac{1}{C}$ respectively of parallel $R-C$ element \cite{16}. Hence, single peak in impedance plot implies that the resistance of grain boundary ($R_{gb}$) is much larger than the grain internal resistance ($R_{g}$) and therefore the impedance plot emphasizes the one with the largest resistance. However, two simultaneous peak in $M''$ indicates comparable value of the capacitance of grain ($C_{g}$) and grain boundary ($C_{gb}$), although $C_{g}$ is lower than $C_{gb}$ according to the magnitude of the peak. 
The frequency of the relaxation peak in both impedance and modulus spectra is inversely proportional to the product of $R$ and $C$ ($\tau$ = $\frac{1}{2\pi f_{P}}$ = $RC$). Hence, above 200 $^{\circ}$C, the position of the high frequency peak in $M''$ does not change but its magnitude changes which indicates that $R_{g}$ decreases the same amount by which $C_{g}$ is increased. 

\subsection{\label{sec:level10}Electrical analysis of 15LN-85CF}

\subsubsection{\label{sec:level11} Impedance spectroscopy}
\begin{figure*}[htbp]
	\centering
	\includegraphics[width=0.8\linewidth]{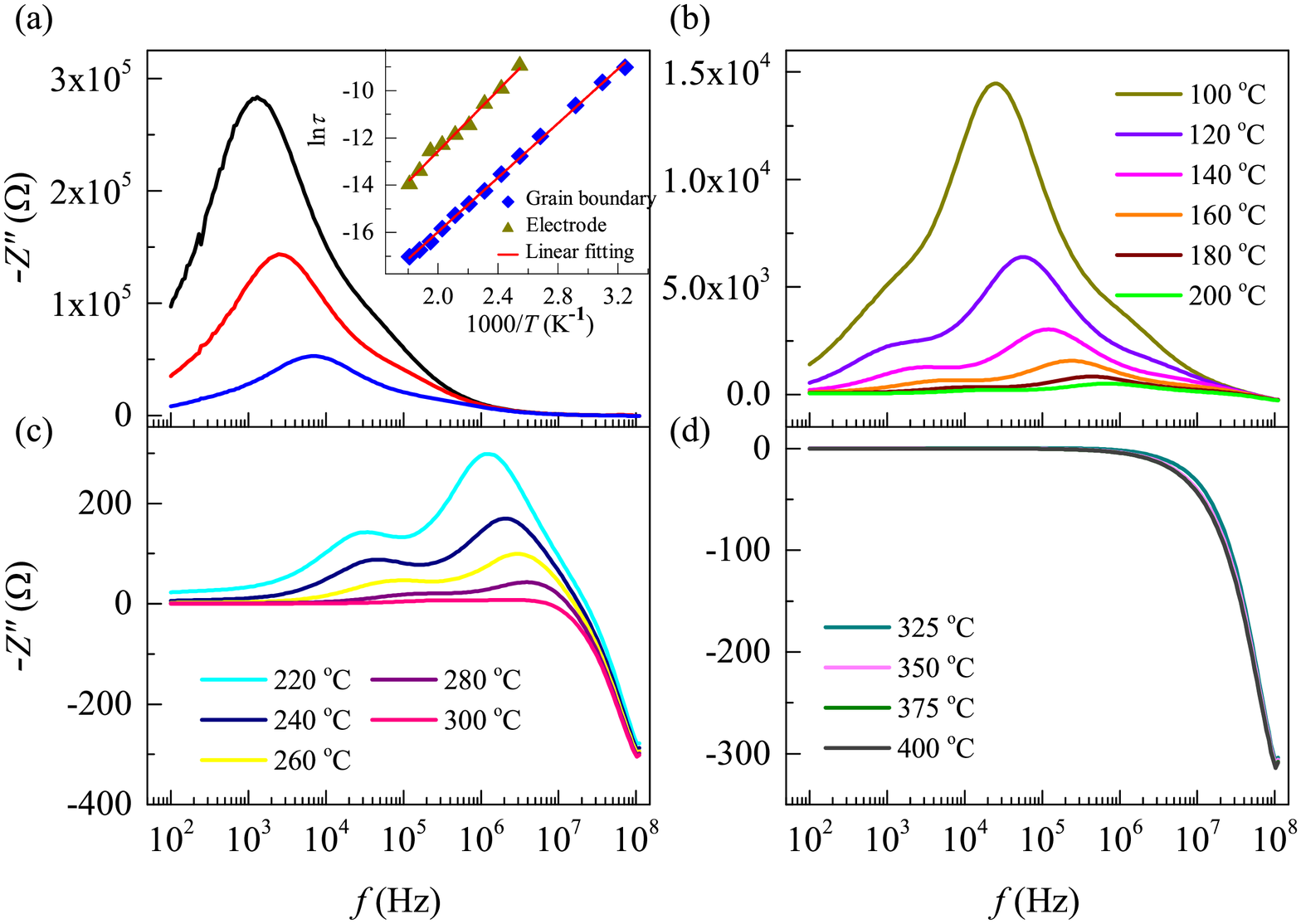}
	\small{\caption{Imaginary part of impedance of the composite 15LN-85CF in the temperature ranges: (a) 35 -- 75 $^\circ$C, (b) 100 -- 200 $^\circ$C, (c) 220 -- 300 $^\circ$C and (d) 325-- 400 $^\circ$C. Inset of (a) shows the linear variation of ln$\tau$ with inverse of temperature for the grain boundary and electrode. \label{fig:15Z2}}}
\end{figure*}

Figure~\ref{fig:15Z2} shows the variation of imaginary part of impedance ($Z''$) with frequency for 15LN-85CF. A strong peak at low frequency and a weak peak around 10$^{5}$ Hz are observed. The magnitude of the strong peak reduces consistently with enhanced temperature. Another weak peak appears at $\sim$ 10$^{3}$ Hz at $\geq$ 100 $^{\circ}$C which was not present in case of pure CFO. The two weak peaks at low and high frequency are related to the relaxation of the accumulated charges inside the grains and at the junction of electrodes and  the surface of the materials, respectively. The strong peak at the intermediate frequency corresponds to the grain boundary effect 
The activation energies as calculated from the grain boundary ($E_{a,gb}$) and electrode ($E_{a,e}$) relaxation are 0.495 and 0.555 eV, respectively using Arrhenius law as shown in the inset of Figure~\ref{fig:15Z2} (a). 
The relaxation peak related to grain effect slowly disappears above 160 $^{\circ}$C while the peak of electrode polarization becomes stronger and more prominent with temperature. Similar to pure CFO, $Z''$ exhibits negative to positive transition at high frequency around 60 MHz at 35 $^{\circ}$C 
Therefore the crossover from capacitive to inductive reactance occurs both in pure CFO and in the composites and the frequency at which the crossover occurs can be tuned by LNO content. The two peaks vanish at 300 $^{\circ}$C and  $Z''$ shows frequency independent behaviour 
upto 10$^{7}$ Hz and then falls to positive values. It shows the inductive nature is more prominent in composite materials compared to pure CFO. 

\subsubsection{\label{sec:level12} Modulus spectroscopy}
\begin{figure}[htbp]
	\centering
	\includegraphics[width=0.95\linewidth]{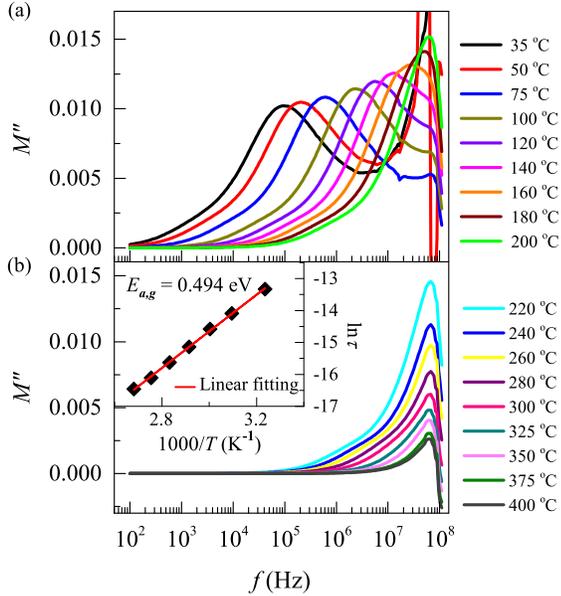}
	\small{\caption{Electric modulus of the composite 15LN-85CF: real part of modulus in the temperature range (a) 35 -- 200 $^\circ$C, (b) 225 -- 400 $^\circ$C and imaginary part of modulus for (a) 35 -- 200 $^\circ$C, (b) 225 -- 400 $^\circ$C. \label{fig:15M}}}
\end{figure}

Figure~\ref{fig:15M} shows the frequency dispersion of the imaginary part ($M''$) of electric modulus of the composite with 15 \% LNO. 
$M''$ shows one clear peak representing the grain relaxation and a point of inflection around 10$^{4}$ Hz. Clearly the electrode peak is completely suppressed in the modulus spectrum ($M''$ vs.~$f$) and the relaxation peak of the grain boundary is also not clear as it was for pure CFO.


Hence considering equivalent circuit for the composite, an additional parallel $R$ - $C$ element should be added to account for the electrode polarization. It implies that the resistance ($R_{gb}$) of the grain boundary are much higher than that of grain ($R_{g}$ and $C_{g}$) and the electrode ($R_{e}$). On the other hand, the electrode is associated with the highest capacitance which is completely suppressed in the modulus spectroscopy. 
The activation energy ($E_{a,g}$) as calculated in the inset of Figure~\ref{fig:15M} (d) is 0.494 eV. It means the activation energy of the grain boundary is reduced by 0.2 eV whereas there is not much affect on $E_{a,g}$. This may be because while adding LNO in the composite, the barrier height at the grain boundary decreases which helps the carriers to easily cross the barrier leading to reduction in the $E_{a,gb}$ significantly whereas $E_{a,g}$ almost remains same.

\subsection{\label{sec:level13} Comparison between impedance and modulus spectroscopy of pure CFO and 15LN-85CF}

\begin{figure}[htbp]
	\centering
	\includegraphics[width=1\linewidth]{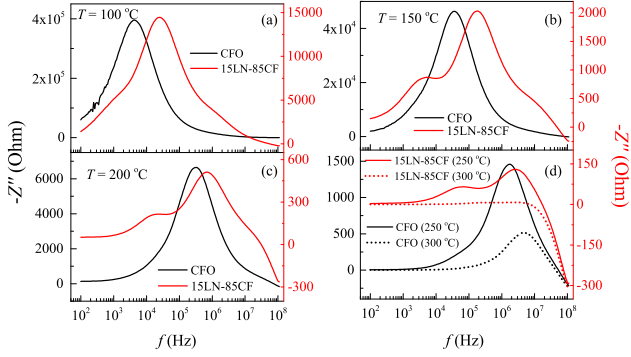}
	\small{\caption{Comparison of the impedance spectroscopy ($Z''$ vs.~$f$) of pure CFO and the composite with 15 \% LNO at temperatures (a) 100 $^{\circ}$C, (b) 150 $^{\circ}$C, (c) 200 $^{\circ}$C, (d) 250 and 300 $^{\circ}$C. \label{fig:cmp}}}
\end{figure}

\begin{figure}[htbp]
	\centering
	\includegraphics[width=1\linewidth]{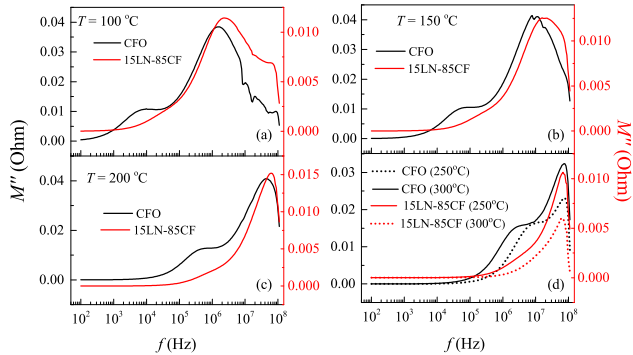}
	\small{\caption{Comparison of the modulus spectroscopy ($M''$ vs.~$f$) of pure CFO and the composite with 15 \% LNO at temperatures (a) 100 $^{\circ}$C, (b) 150 $^{\circ}$C, (c) 200 $^{\circ}$C, (d) 250 and 300 $^{\circ}$C.\label{fig:cmp1}}}
\end{figure}

To get a comparative view between pure CFO and the composite with 15 \% LNO, the impedance and modulus spectroscopy are plotted in \Fref{fig:cmp} and \ref{fig:cmp1} respectively at five temperatures 100, 150, 200, 250 and 300 $^{\circ}$C. 

As evident from the \Fref{fig:cmp}, on adding 15 \% LNO, the grain boundary resistance of pure CFO is reduced by one order of magnitude at all temperatures which makes the peak of grain relaxation comparable and visible in the impedance spectroscopy. However, \Fref{fig:cmp1} indicates that the capacitance of the composite is enhanced compared to pure CFO. The capacitance is formed when a thin layer of CFO is trapped and embedded on both sides by conducting LNO. Therefore, with adding LNO, the chance of capacitance formation is increased while the increasing number of the free charge carriers from LNO reduces the resistances.

\subsection{\label{sec:level14} Dielectric properties of pure CFO}

The dielectric constant ($\epsilon^{*}$) is calculated using the following expression:

\begin{equation}
\varepsilon^{*} = \varepsilon'+ j\varepsilon''= \frac{1}{j\omega C_{0}Z^{*}}
\label{eq.4}
\end{equation}   

The real ($\epsilon'$) and imaginary ($\epsilon''$) parts of the dielectric constant can be expressed as:

\begin{equation}
\varepsilon' = \frac{1}{\omega C_{0}}\Big[\frac{Z''}{Z'^{2}+Z''^{2}}\Big]
\label{eq.5}
\end{equation} 

and

\begin{equation}
\varepsilon'' = \frac{1}{\omega C_{0}}\Big[\frac{Z'}{Z'^{2}+Z''^{2}}\Big]
\label{eq.6}
\end{equation} 

\begin{figure}[htbp]
	\centering
	\includegraphics[width=1\linewidth]{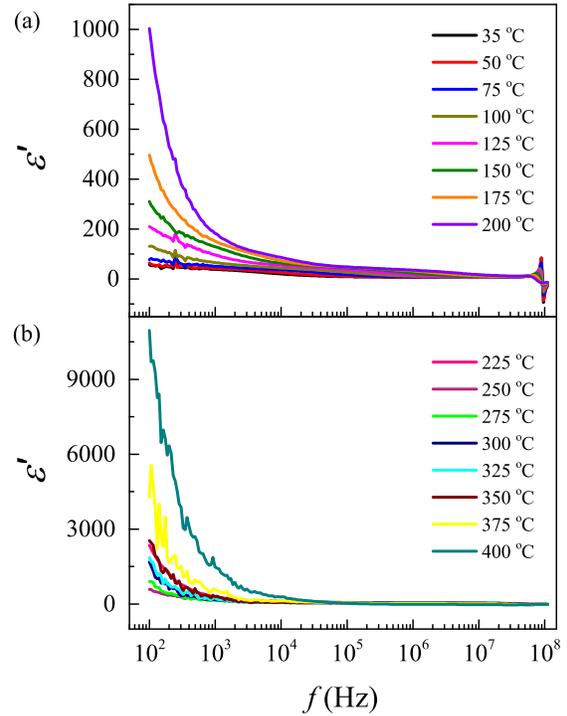}
	\small{\caption{Dielectric properties of pure CFO: real part of the dielectric constant at (a) 35 -- 200 $^\circ$C, (b) 225 -- 400 $^\circ$C. 
			\label{fig:E}}}
\end{figure}

The frequency dependence of the real part of dielectric constant ($\varepsilon'$) of pure CFO is shown in Figure~\ref{fig:E} (a) for $T \leq$ 200 $^{\circ}$C and (b) for 225 $^{\circ}$C $\leq T \leq$ 400 $^{\circ}$C. It displays that $\varepsilon'$ increases with temperature and the increment is faster and more pronounced at low frequency region. In CFO, two types of charge carriers are present, the p-type charge carriers due to hole hopping between Co$^{+2}$$\leftrightarrow$Co$^{+3}$ and n-type charge carriers due to electron hopping between $Fe^{+2}$$\leftrightarrow$Fe$^{+3}$ \cite{29}. 
When the temperature is increased, high thermal energy is available to these charge carriers and therefore the mobility and the hopping rate are enhanced which gives rise to higher dielectric constant. Usually in ferrites, the dielectric constant  arises due to four basic types of polarizations: electronic, atomic, orientational and interfacial or space charge polarization \cite{29}. The interfacial and orientational polarization play major roles at low frequency whereas the dielectric constant at high frequency  is attributed to mainly atomic and electronic polarization. The interfacial and orientational polarization involve the inelastic movement of charge carriers. The orientation of dipoles is a rotational process hence it involves the energy required to overcome the resistance due to thermal agitation and the inertia of the surrounding molecules. Hence the orientation and interfacial polarizations are intermolecular processes and strongly depend on temperature \cite{18}. This explains the rapid increase of $\varepsilon'$ with rising temperature at low frequency. But in Atomic or electronic polarization, the elastic displacement of the electron clouds or the ions takes place and the restoring force against the displacement only slightly depends on temperature, giving rise to negligible temperature dependence of $\varepsilon'$ at high frequency.

The frequency dispersion of $\varepsilon'$ reveals that $\varepsilon'$ has large value at low frequency and it decreases with increasing frequency which is normal dielectric behaviour of ferrites \cite{29,30,31}. This behaviour can be explained by Koops's phenomenological theory based on Maxwell-Wagner type two layers model \cite{32}. According to this model, we can describe the polycrystalline dielectric material as an inhomogeneous medium consisting of well-conducting grains separated by poorly conducting grain boundaries. 
At low frequency, the contribution of grain boundaries to dielectric constant is more effective and hence dominates over the grains. In ferrites, the polarization originates from the same mechanism as the electrical conduction \cite{33,30}. The ferrites are dipolar due to the presence of Fe$^{+3}$ and the minority Fe$^{+2}$ ions \cite{34}. The alignment of the dipoles Fe$^{+2}$-Fe$^{+3}$ along the direction of the applied field occurs via the rotation which can be pictured as the hopping of charge carriers between the ions Fe$^{+2} \leftrightarrow$Fe$^{+3}$. As the resistance of the grain boundary is very high, the electrons after reaching the grain boundary by hopping between the ions, pile up there and leads to space charge polarization which causes very high value of dielectric constant at lower frequency \cite{30,31,35}. The dielectric constant depends on the number of ions, the higher the concentration of ions (Fe$^{+2}$ and Fe$^{+3}$) the larger the dielectric constant should be \cite{29}. As the frequency of the applied electric field is increased, the hopping frequency of the electrons can not follow the fast variation of the applied field and lags behind it beyond a certain critical frequency \cite{36,37}. This leads to reduced polarization and hence decrease in dielectric constant with increasing frequency and then a constant $\varepsilon'$.

\begin{figure}[htbp]
	\centering
	\includegraphics[width=1\linewidth]{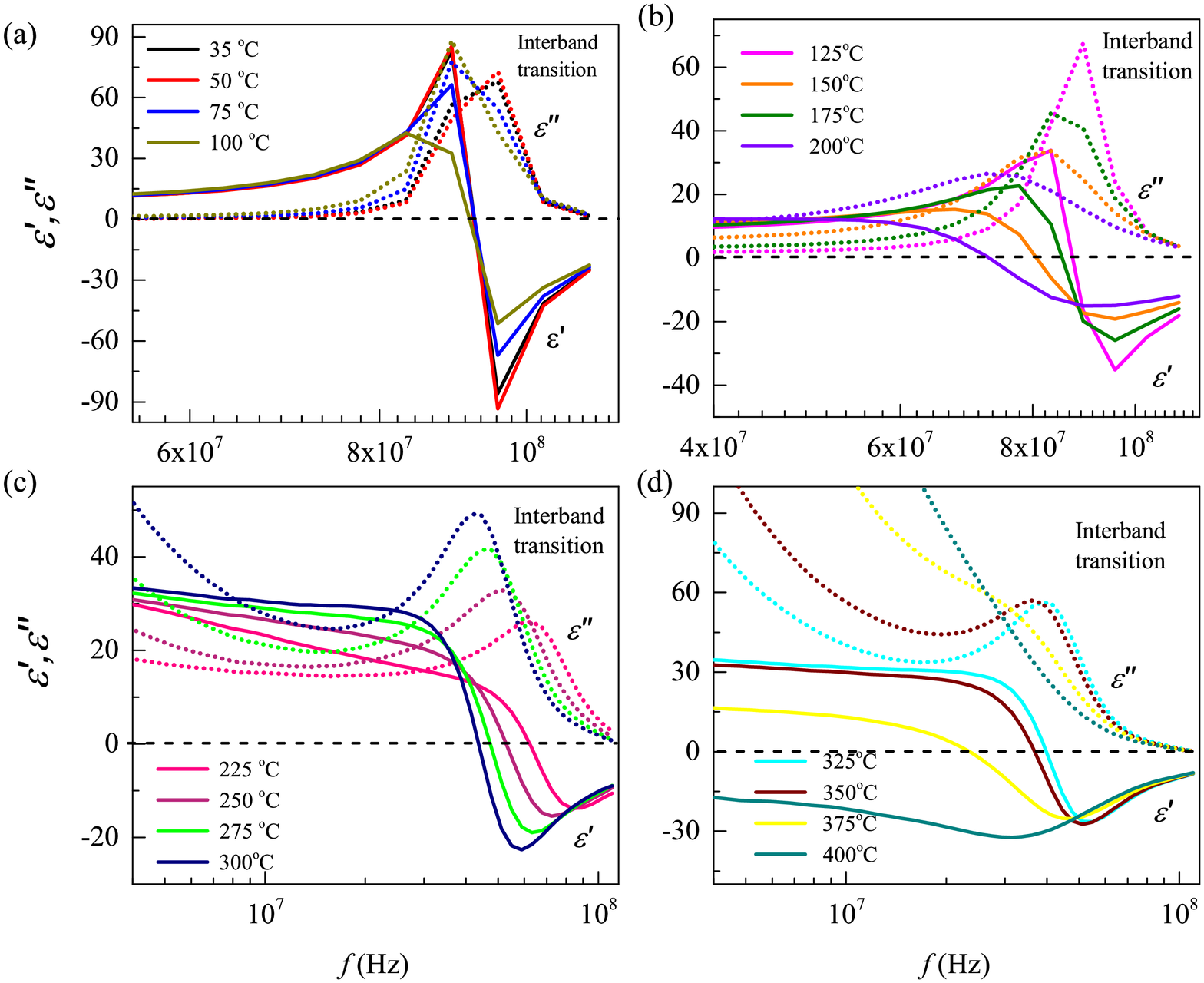}
	\small{\caption{The real and imaginary part of dielectric constant of pure CFO at high frequency ($f >$ 10 MHz)  at four different temperature regimes: (a) 35 -- 100 $^\circ$C, (b) 125 -- 200 $^\circ$C, (c) 225 -- 300 $^\circ$C and (d) 325 -- 400 $^\circ$C; the negative value in real part associated with a peak in the imaginary part implies the interband transition. \label{fig:EP}}}
\end{figure}

At high frequency region ($\sim$ 90 MHz at 35 $^{\circ}$C), $\varepsilon'$ shows an unique nature where it slowly goes up, reaches a maximum then suddenly comes down to negative value and then goes back to $\sim$ zero at all temperatures as shown in Figure~\ref{fig:EP}. With increasing temperature, the frequency ($f_{0}$) at which $\varepsilon'$ switches to negative value slowly shifts to lower frequency with the reduced amplitude of $\varepsilon'$ in both positive and negative parts. The corresponding imaginary part of dielectric constant ($\varepsilon''$) is also shown in the same figure which manifests a peak around $f_{0}$. 
This behaviour results from the interband transition where the high energy photons excite the bound electrons from low lying bands to conduction bands above the Fermi energy \cite{38,39,40}. The shifting of $\varepsilon'$ to negative value is also observed in CNT/polymer composites \cite{39,40} or in Ag/Al$_{2}$O$_{3}$ \cite{25}, Fe/Al$_{2}$O$_{3}$ composites \cite{41} or in La doped PbTiO$_{3}$ under uniaxial stress \cite{42}. We have found that $f_{0}$ matches with the frequency where the capacitive to inductive transition occurs in $Z''$. Considering interband transition the contribution of bound electrons to real and imaginar part of dielectric constant can be written as \cite{38}:

\begin{equation}
\varepsilon'= 1+\frac{\omega_{p}^{2}(\omega_{0}^2-\omega^{2})}{(\omega_{0}^2-\omega^{2})^2+(\gamma \omega)^2},  
\label{eq.7}
\end{equation}

\begin{equation}
\varepsilon'' = \frac{\gamma\omega_{p}^{2}\omega}{(\omega_{0}^2-\omega^{2})^2+(\gamma \omega)^2}
\label{eq.8}
\end{equation} 

\begin{figure}[htbp]
	\centering
	\includegraphics[width=1\linewidth]{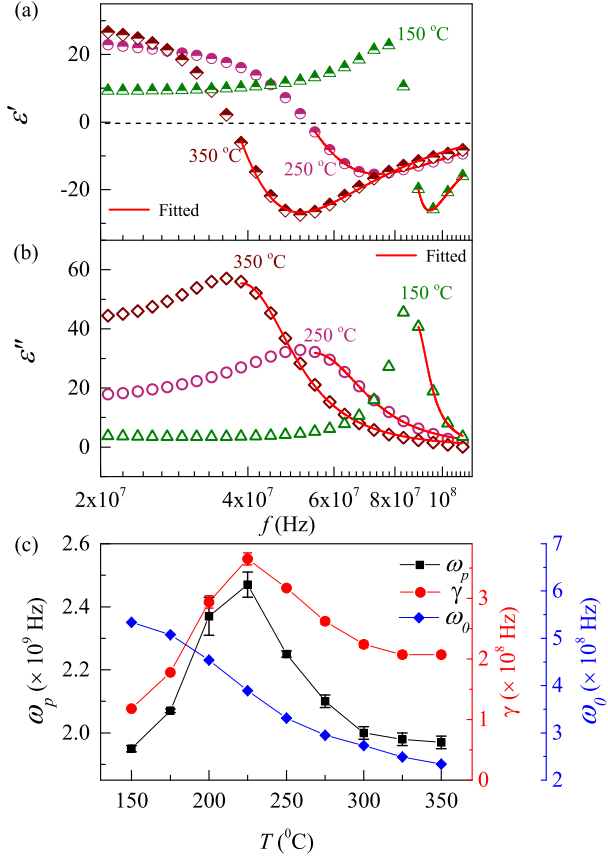}
	\small{\caption{For pure CFO, (a) the real and (b) imaginary parts fitted with Eqs.~\ref{eq.5.16} and \ref{eq.5.17} respectively, shown at three temperatures 150, 250 and 350 $^\circ$C. (c) The variations of the fitted parameters with temperature. \label{fig:wp}}}
\end{figure}
where $\omega_{p}$ is the analogy to the Plasma frequency in Drude model given by $\omega_{p}$ = $\sqrt{\textit{ne}^{2}/\textit{m}^{*}\varepsilon_{0}}$. Here $n$ is the density and $m^{*}$ is the effective mass of bound electrons. $\gamma$ is the damping constant. $\omega_{0}$ is the defined as $\sqrt{\alpha/m^{*}}$ where $\alpha$ is the spring constant of the potential to keep the electrons in space. Our data of $\varepsilon'$ and $\varepsilon''$ follow the equations of interband transitions and are fitted from 150 $^{\circ}$C to 350 $^{\circ}$C. The fitting is shown in Fig.~\ref{fig:wp} (a) and (b) for $\varepsilon'$ and $\varepsilon''$ respectively for three temperatures (150, 250 and 350 $^{\circ}$C) and the fitted parameters are presented in Fig.~\ref{fig:wp} (c) which shows that both $\omega_{p}$ and the damping constant first increase upto 225 $^{\circ}$C and then decrease with increasing temperature whereas the $\omega_{0}$ decreases consistently. Similar temperature variation of plasma frequency is observed in polycrysttaline thin film of Gold \cite{43}. $\omega_{p}$ depends on number density and effective mass of electrons. As the temperature increases the electron density is reduced due to the thermal expansion of the volume of crystals. The thermal expansion  coefficient of CFO is 14.9 $\times$ 10$^{6}$ K$^{-1}$ \cite{44}. Now depending on the curvature of the band edge, the effective mass increases or decreases with temperature.

Therefore, the initial increase of $\omega_{p}$ indicates a larger decrease in effective mass compared to carrier density \cite{45,43} while above 225 $^{\circ}$C, the reduced number density dominates the change in $\omega_{p}$. 
The change in $\omega_{0}$ can be understood if we consider it is analogous to the natural frequency of free electron system in Drude model. With increasing thermal energy the spring constant of the potential is reduced, so $\omega_{0}$  decreases continuously with temperature. In bulk CFO, the value of plasma frequency is much higher ($\omega_{p} \sim$ 10$^{15}$ Hz) \cite{46}. As the particle size decreases, the effective mass increases because the curvature of the uppermost valence band and lowest conduction band is reduced \cite{47}. Also it is theoretically predicted that when the electrons are confined in a thin wire, the effective mass is increased by several orders and the average electron density is reduced \cite{48}. Therefore the exceptional reduction in $\omega_{p}$ could be attributed to the nanocrystalline structure of the present CFO, since $\omega_{p}$ depends on the size and shape of the nanoparticles.
\begin{figure}[htbp]
	\centering
	\includegraphics[width=1\linewidth]{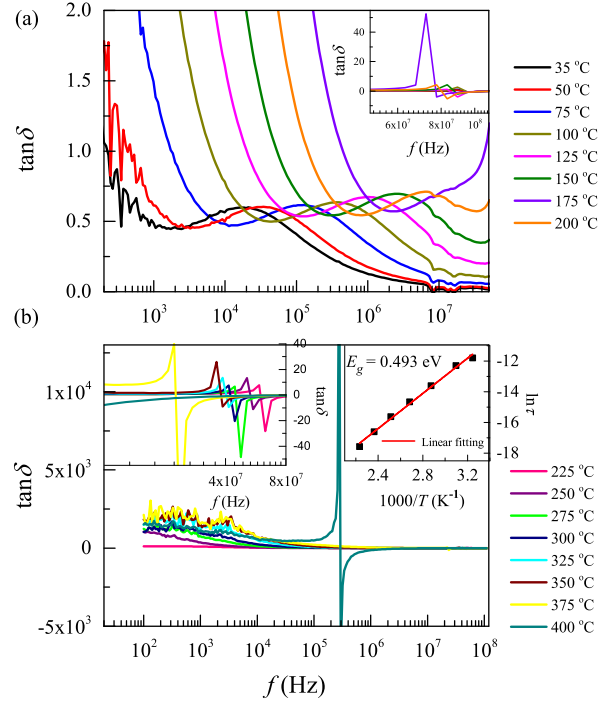}
	\small{\caption{Dielectric loss at (c) 35 -- 200 $^\circ$C, (d) 225 -- 400 $^\circ$C. The inset of (c) and left inset of (d) shows the negative value of the tangent loss at high frequency. The right inset of (d) represents the resonance peak in tan$\delta$ follows thermally activated Arrhenius equation.\label{fig:tan}}}
\end{figure}
All kinds of polarization experience some inertia which try to prevent the displacement of electrons, movement of charge carriers or orientation of dipoles and therefore involve some energy loss which must be consumed to overcome the resistance of the inertia. This energy loss is called dielectric loss or tangent loss \cite{18} defined as the ratio of imaginary part and real part of dielectric constant ($\varepsilon''/\varepsilon'$). The tangent loss of pure CFO is shown in Figure~\ref{fig:tan} (a) ($T \leq$ 200 $^{\circ}$C) and (b) (200 $^{\circ}$C $< T \leq$ 400 $^{o}$C). The dielectric loss exhibits a peak around 30 KHz at 35 $^{\circ}$C which moves slowly to high frequency with raising temperature. When the frequency of the applied electric field is close to the hopping frequency of the charge carriers, maximum energy is absorbed by the carriers and a peak is observed in the loss. This phenomenon is called resonance which is observed earlier in nanostructured  CFO \cite{9} or CFO thick film \cite{49} and other ferrites \cite{50}. This resonance peak follows Arrhenius equation (Eq.~\ref{eq.1}) which indicates thermally activated process with activation energy 0.493 eV (right inset of Figure~\ref{fig:E} (d)). At high frequency, tan$\delta$ shows negative value as $\varepsilon'$ at the same frequency indicating interband transition as shown in the inset of Figure~\ref{fig:E} (c) and left inset of (d). 

\subsection{\label{sec:level15} Dielectric properties of 15LN-85CF}

\begin{figure}[htbp]
	\centering
	\includegraphics[width=1\linewidth]{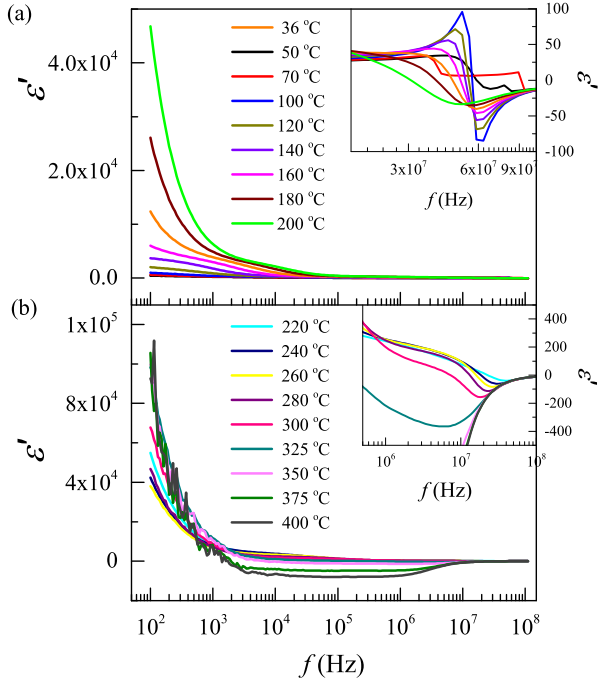}
	\small{\caption{Real part of the dielectric constant of the composite 15LN-85CF at (a) 35 -- 200 $^\circ$C, (b) 220 -- 400 $^\circ$C. 
			\label{fig:15E}}}
\end{figure}

The real part of dielectric constant ($\varepsilon'$) for the composite 15LN-85CF is shown in Fig.~\ref{fig:15E} (a) for $T \leq$ 200 $^{\circ}$C and (b) for 220 $^{\circ}$C $\leq T \leq$ 400 $^{\circ}$C. $\varepsilon'$ decreases with increasing frequency in similar manner as pure CFO and at high frequency the interband transition effect causes similar anomalous behaviour as shown in the inset of Figure~\ref{fig:15E} (a) and (b). Although the composite exhibits negative dielectric constant at lower frequency ($f_{0} \sim$ 60 MHz at 35 $^{\circ}$C) compared to pure CFO. As LNO is added in the composites, the concentration of bound electron is reduced significantly as the electrons start to de-localize and hence $f_{0}$ shifts to lower frequency. 

\begin{figure}[htbp]
	\centering
	\includegraphics[width=1\linewidth]{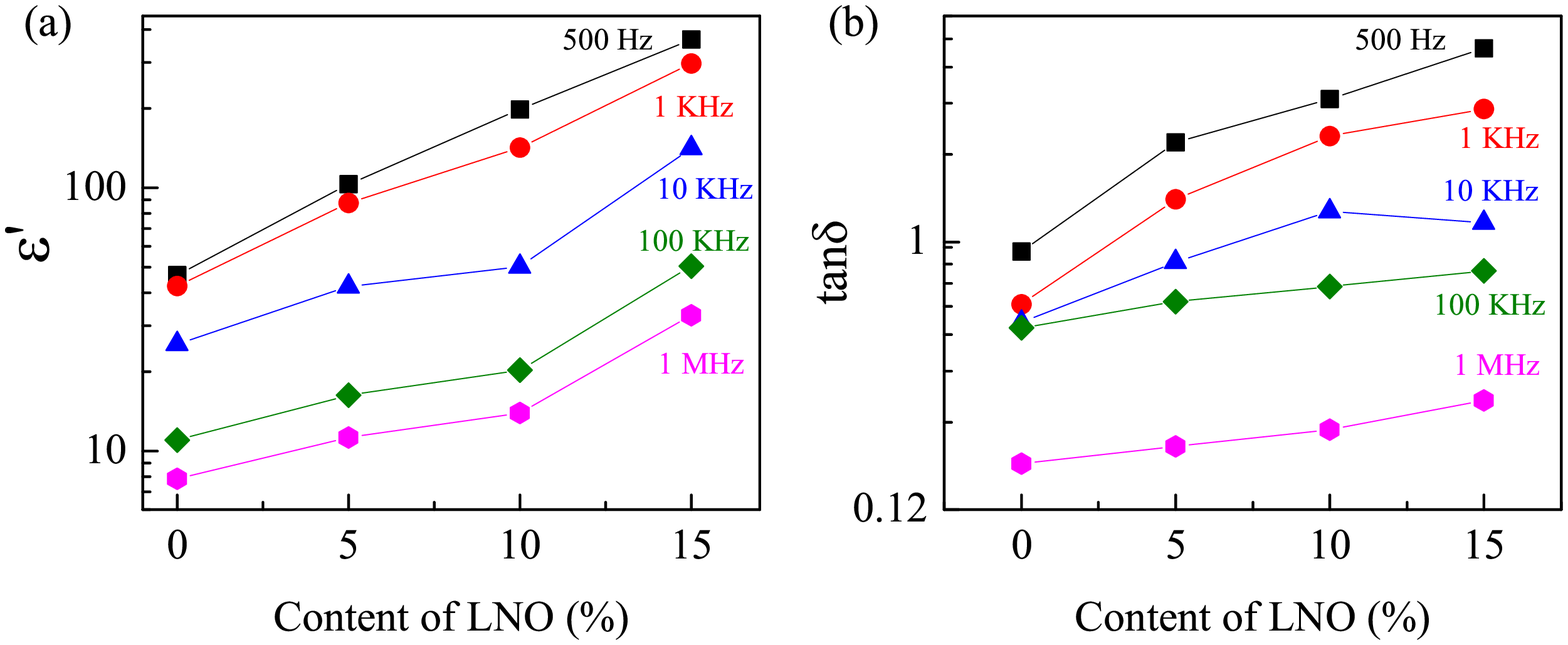}
	\small{\caption{Comparison of the real part of dielectric constant and the tangent loss with varying LNO content at 50 $^\circ$C for different frequencies.  \label{fig:Eall}}}
\end{figure}

For comparison we have plotted $\varepsilon'$ and tan$\delta$ for the series of composites [xLNO + (1-x)CFO, x = 0, 0.05, 0.10 and 0.15] as a function of LNO content (\%) at 50 $^{\circ}$C in Fig.~\ref{fig:Eall} (a) and (b) respectively. Both the dielectric constant and dielectric loss systematically increase with increasing LNO content, although increment in $\varepsilon'$ is much faster than tan$\delta$. For example, at 1 KHz the dielectric constant increases seven times while the loss increases four times in magnitude in the composite with 15 \% LNO compared to pure CFO. Therefore 15LN-85CF is a better material in the application of radio and microelectronic communications compared to pure CFO. The significant increase in dielectric constant of the composite can be explained by Maxwell-Wagner-Sillar type polarization effect between the interfaces of LNO and CFO. As LNO content is increased, more number of charge carriers are accumulated at the LNO/CFO  interfaces due to their different dielectric constants and different electrical conductivities which results in increasing polarization and as a result increase in dielectric constant \cite{51,52}. Similar phenomena are reported in other composites of insulating matrix with metallic filler content close to their percolation threshold such as in Ag/Al$_{2}$O$_{3}$ composites \cite{25} or in polymer-based nanocomposites \cite{53, 52, 54}. 

\begin{figure}[htbp]
	\centering
	\includegraphics[width=1\linewidth]{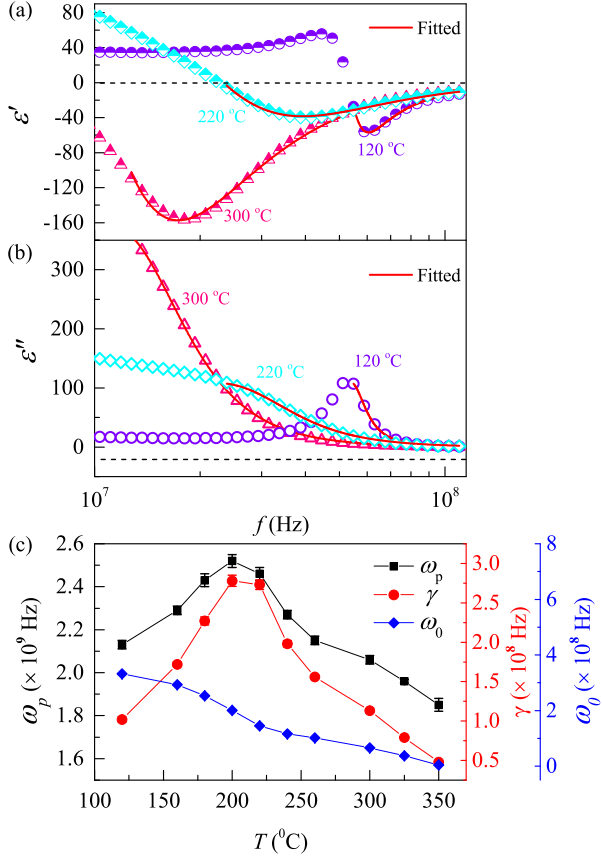}
	\small{\caption{For the composite with 15 \% LNO, (a) the real and (b) imaginary parts fitted with Eqs.~\ref{eq.5.16} and \ref{eq.5.17} respectively, shown at three temperatures 120, 220 and 300 $^\circ$C. (c) The variations of the fitted parameters with temperature.\label{fig:15wp}}}
\end{figure}

The dielectric constant of the composite 15LN-85CF also follows the behaviour of the interband transition  (Eqs.~\ref{eq.7} and \ref{eq.8}) at high frequency region (Fig.~\ref{fig:15wp} (a) and (b)). The fitted parameters as shown in Fig.~\ref{fig:15wp} (c) follow same behaviour with temperature as pure CFO indicating reduced effective mass and reduced carrier density are the dominating mechanisms below and above 220  $^{\circ}$C, respectively. 

\subsection{\label{sec:level16} AC conductivity of pure CFO}

\begin{figure}[htbp]
	\centering
	\includegraphics[width=1\linewidth]{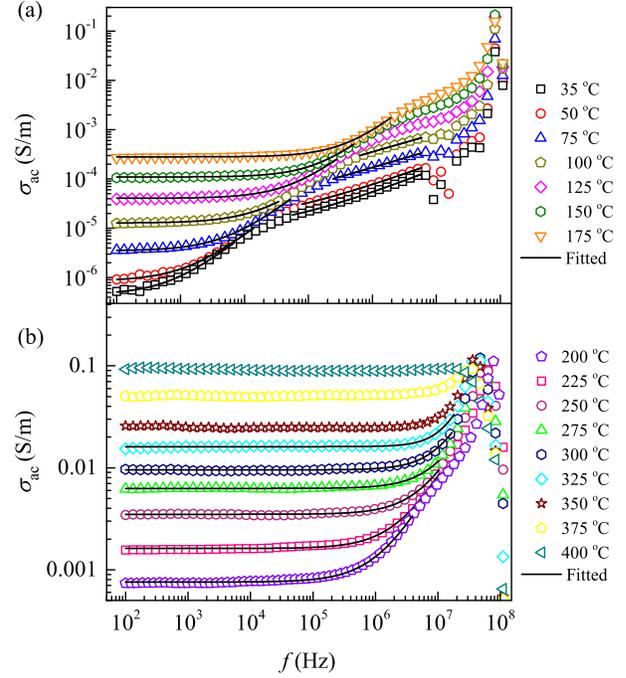}
	\small{\caption{Real part of ac conductivity of pure CFO for temperature (a) 35 -- 175 $^\circ$C, (b) 200 -- 400 $^\circ$C \label{fig:ac1}}}
\end{figure}

In order to understand the transport mechanism, the real part of ac conductivity ($\sigma_{ac}$) is studied as a function of frequency for pure CFO as shown in Fig.~\ref{fig:ac1}. 
At temperature below 175 $^{\circ}$C, two  frequency dependent plateau regions are observed with different increasing rate at low and high frequency sides which are separated by a small hump around $\sim$ 10$^{4}$ Hz at 35 $^{\circ}$C. As the temperature increases 
the high frequency plateau slowly becomes more dispersive 
whereas the low frequency plateau 
takes a form of frequency-independent dc conductivity which becomes dominating above 300 $^{\circ}$C over a large frequency range. The low frequency plateau and the dc conductivity is associated with grain boundary which is more effective with higher resistance at low frequency region. The plateau at high frequency side corresponds to the conductivity inside the grains which becomes more active with increasing frequency  \cite{30,55}. At very high frequency ($\sim$ 90 MHz at 35 $^{\circ}$C), the ac conductivity reaches a maximum value and then starts to decrease with further increasing frequency. This occurs at the same frequency where $Z''$ becomes positive and hence indicates the free charge conduction. 

The frequency dispersion of ac conductivity can be explained by Jump relaxation model proposed by Funke \cite{56} and short range hopping. According to jump relaxation model, at low frequency the mobile ions hop successfully to its neighbouring vacant sites due to long available time period. Many such consecutive successful jumps give rise to long-range translational motion of the ions which is responsible for the dc conductivity plateau at low frequency for higher temperature ($>$ 75 $^{\circ}$C).



At higher frequency ($>$ 10$^{4}$ Hz), two types of hopping processes may occur: (i) successful hopping where the ion jumps to the neighbouring site, becomes relaxed with respect to its position and stays at the new site, (ii) the ion comes back to its initial position after jumping to the neighbouring site that is unsuccessful hopping also called correlated forward-backward-forward hopping \cite{57,58,10}. Depending on the increasing ratio of successful to unsuccessful hopping, the conductivity becomes dispersive as the frequency increases. The frequency dependent conductivity follows Jonscher's universal power law which is given by the expression \cite{17}:

\begin{equation}
\sigma_{ac}= \sigma_{dc}+A\omega^{n} ,   
\label{eq.9}
\end{equation}

where $\sigma_{ac}$ is the total conductivity, $\sigma_{dc}$ is the frequency independent conductivity, $A$ is the coefficient and $n$ is the exponent. These parameters depend on the inherent properties of the material \cite{59}. The two plateau regimes are fitted separately with Eq.~\ref{eq.9} as shown in Fig.~\ref{fig:ac1} for different temperatures. 
The dc conductivity ($\sigma_{dc}$) follows Arrhenius type thermal activation process:

\begin{equation}
\sigma_{dc}= \sigma_{0}exp(-E_{a}/k_{B}T) ,   
\label{eq.5.19}
\end{equation}
where $\sigma_{0}$ is a constant, $E_{a}$ is the activation energy and $k_{B}$ is the Boltzmann constant. 

\begin{figure}[htbp]
	\centering
	\includegraphics[width=1\linewidth]{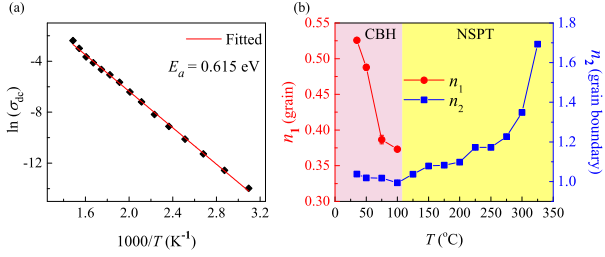}
	\small{\caption{(a) The dc conductivity following thermally activated Arrhenius equation, (b) the temperature variation of the exponenets $n_{1}$ and $n_{2}$ corresponding to grain and grain boundary, respectively. \label{fig:acpara}}}
\end{figure}

The activation energy as calculated from the fitting of ln($\sigma_{dc}$) vs. 1000/$T$ (Fig.~\ref{fig:acpara} (a)) is 0.615 eV which is close to that obtained from imaginary impedance for grain boundary effect. The parameters $n_{1}$ and $n_{2}$ denote the exponent associated with the dispersion of grain and grain boundary, respectively and their variation with temperature is shown in Fig.~\ref{fig:acpara} (b). It shows that $n_{1}$ varies between 0 and 1 whereas $n_{2}$ varies between 1 and 2. Therefore $n_{1}$ indicates the short range hopping inside the grains and $n_{2}$ denotes the localized or reorientation hopping along the grain boundary \cite{56,57}. The localized hopping is the motion of carriers between two charge defects in back and forth direction and re-orientational hopping is related to the orientation of permanent or induced dipoles \cite{57}. Both $n_{1}$ and $n_{2}$ decreases upto 100 $^{\circ}$C and then $n_{2}$ increases with increasing temperature. This temperature dependence of the exponents can be explained based on these two models, correlated barrier hopping model (CBH) \cite{60} and non-overlapping small polaron tunneling model (NSPT) \cite{61,62}. In case of CBH model, the conduction occurs when the electron hops between two charged defect sites over a potential barrier. Each site is associated with the Coulomb potential well and the two wells overlap at a distance $R$. As a result the barrier height decreases from $W_{M}$ to $W$. The exponent $n$ can be expressed by this formula \cite{62}:

\begin{equation}
n = 1-\frac{6k_{B}T}{W_{M}-k_{B}Tln(\frac{1}{\omega\tau_{0}})},   
\label{eq.5.20}
\end{equation}

where $W_{M}$ is the maximum barrier height and $\tau_{0}$ is the characteristic relaxation time. With increasing temperature, the parameter $W_{M}$ decreases and hence it leads to reduction in $n$.

Considering NSPT model, a polaron in a distorted lattice is an electron which is kept inside a potential well created by the lattice deformation. When the polaron is confined in very small volume of the order of one unit cell or less, it is called small polaron. The effective mass of the small polaron is very high and therefore its kinetic energy in potential well is almost negligible. The conduction in this model occurs when the small polaron tunnels from one localized site to the other. The exponent $n$ is given by the expression \cite{62}:

\begin{equation}
n = 1-\frac{4}{ln(\frac{1}{\omega\tau_{0}})-\frac{W_{H}}{k_{B}T}},   
\label{eq.5.21}
\end{equation}

where $W_{H}$ is polaron hopping energy and $\tau_{0}$ is the characteristic relaxation time.

In this model, the exponent increases with increasing temperature.

Therefore in case of pure CFO, for both grain and grain boundary the conduction follows CBH model below 100 $^\circ$C. And above that temperature small polaron assisted reorientational hopping takes place around the grain boundary.


\subsection{\label{sec:level17}AC conductivity of 15LN-85CF}

\begin{figure}[htbp]
	\centering
	\includegraphics[width=1\linewidth]{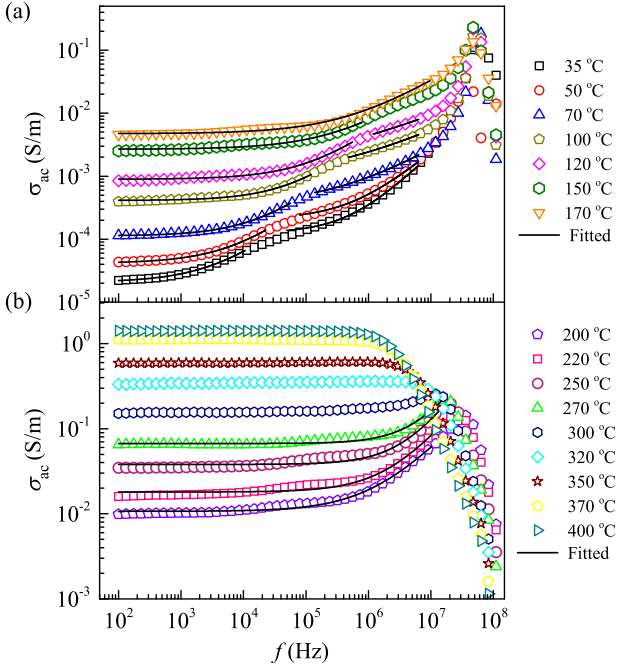}
	\small{\caption{Real part of ac conductivity of the composite with 15 \% LNO for temperature (a) 35 -- 170 $^\circ$C, (b) 200 -- 400 $^\circ$C\label{fig:15ac}}}
\end{figure}

\begin{figure}[htbp]
	\centering
	\includegraphics[width=1\linewidth]{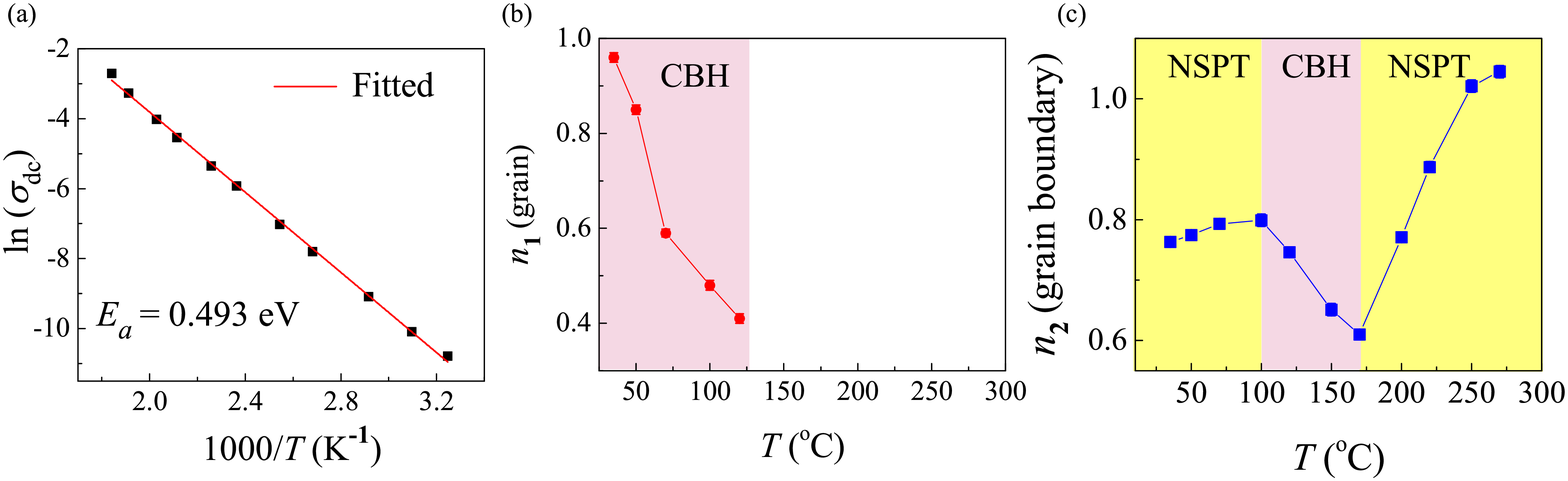}
	\small{\caption{For the composite with 15 \% LNO, (a) the dc conductivity following thermally activated Arrhenius equation, (b) the temperature variation of the exponenets (a) $n_{1}$ and (b) $n_{2}$ corresponding to grain and grain boundary, respectively. \label{fig:15acpara}}}
\end{figure}

The real part of ac conductivity for the composite with 15 \% LNO is presented in Fig.~\ref{fig:15ac} in the temperature range 35--170 $^{\circ}$C and 200--400 $^{\circ}$C. The ac conductivity is increased more than one order of magnitude in the composites as expected due to the presence of conductive LNO. Similar to pure CFO, $\sigma_{ac}$ shows two frequency dependent plateaus signifying grain and grain boundary contributions. 
The free electron conduction is also present in the composites and it occurs
comparatively lower frequency ($\sim$ 60 MHz). The grain and grain boundary contributions are fitted separately with the Jonscher's power law (Eq.~\ref{eq.5.18}) and the dc conductivity following Arrhenius law (Fig.~\ref{fig:15acpara} (a)) provides an activation energy 0.493 eV. The values of the fitted parameters are  shown in Fig.~\ref{fig:15acpara} (b). The exponent $n_{1}$  representing grain conduction shows similar behaviour as pure CFO. The value varies between 0 and 1 and it follows CBH model. However, the grain boundary conduction indicates short range hopping as the exponent $n_{2}$ varies between 0 and 1 which is different than pure CFO. This result supports our finding of the reduced activation energy of the grain boundary ($E_{a,gb}$) 
from impedance spectroscopy. In pure CFO when $E_{a,gb}$ is high the charges and the induced or permanent dipoles are accumulated around the grain boundary and conduction occurs only via the orientation of the dipoles in the direction of the electric field. On the other hand, on adding LNO, when $E_{a,gb}$ is reduced and it helps the charge carriers to easily hop the barriers which leads to short range hopping across the grain boundary. From the temperature variation of $n_{2}$ it shows that the grain boundary conduction is assisted by small polaron tunneling upto 100 $^\circ$C, then it follows CBH model upto 170  $^\circ$C and above that again small polaron tunneling takes place.

\section{\label{sec:level18}Conclusion}

To conclude, both grain and grain boundary relaxations are observed in impedance and modulus spectroscopy for pure CFO. In case of the composite with 15 \% LNO, an additional relaxation appears at the low frequency side ($\sim$ 1 KHz) at temperature $\geq$ 100 $^\circ$C which is ascribed to the electrode polarization. The dielectric constant of both pure CFO and the composite materials manifest a transition from positive to negative value due to the interband transition. The frequency at which interband transition occurs shifts to lower frequency with adding LNO. The dielectric constant of the composite 15LN-85CF is enhanced compared to pure CFO which can be explained by Maxwell-Wagner-Sillar type polarization theory. Therefore, considering both dielectric constant and dielectric loss the composite with 15 \% LNO is better for microelectronics applications than pure CFO. In pure CFO, the conduction occurs via short range hopping inside the grains and reorientational hopping across the grain boundary. While in the composite 15LN-85CF, the activation energy of the grain almost remains unchanged therefore the nature of the conduction process inside grains is same as pure CFO. However, the activation energy of the grain boundary is reduced significantly (by 0.2 eV) and it leads to short range hopping of the charge carriers across the grain boundary.


\section*{References}
\begin{thebibliography}{75}

\bibitem{1} Qi X, Zhou J, Yue Z, Gui Z, Li L and Buddhudu S 2004 {\it Advanced Functional Materials} {\bf 14} 920

\bibitem{2} Vadivel M, Ramesh Babu R, Ramamurthi K and Arivanandhan M 2017 {\it Journal of Physics and Chemistry of Solids} {\bf 102} 1

\bibitem{3} Daniel H, Kumary T. Geetha, Lin L and Lin J G 2006 {\it Phys. Rev. B} {\bf 74} 214504

\bibitem{4} Lin J. J. 1991 {\it Journal of Applied Physics} {\bf 69} 7723

\bibitem{5} Silvestre J, Silvestre N and De Brito J 2015 {\it Journal of Nanomaterials} {\bf 2015} 3

\bibitem{6} Bajpai A and Nigam A K 2004 {\it Phys. Rev. B} {\bf 75} 064403

\bibitem{7} Johnson H G, Bennett S P, Barua R, Lewis L H and Heiman D 2010 {\it Phys. Rev. B} {\bf 82} 085202

\bibitem{8} Patra A, Maity K P, Kamble R B and Prasad V 2018 {\it Journal of Physics: Condensed Matter} {\bf 30} 375701

\bibitem{9} Sharma J, Parashar J, Saxena V K, Bhatnagar D and Sharma K B 2015 {\it Macromolecular Symposia} {\bf 357} 38

\bibitem{10} Chen W, Zhu W, Tan O K and Chen X F 2010 {\it Journal of Applied Physics} {\bf 108} 034101

\bibitem{11} Das A, De S, Bandyopadhyay S, Chatterjee S and Das D 2017 {\it Journal of Alloys and Compounds} {\bf 697} 353

\bibitem{12} Dabas S, Chaudhary P, Kumar M, Shankar S and Thakur O P 2019 {\it Advanced Functional Materials} {\bf 30} 2837

\bibitem{13} Acevedo U, Gaudisson T, L{\'o}pez-Noda R, Ammar S, Nowak S and Valenzuela R 2012 {\it Advanced Functional Materials} {\bf 1} 85

\bibitem{14} Maaz K, Mumtaz A, Hasanain S K and Ceylan A 2007 {\it Advanced Functional Materials} {\bf 308} 289

\bibitem{15} Hodge I M, M.D. Ingram M D and West A R 1976 {\it Advanced Functional Materials} {\bf 74} 125

\bibitem{16} Sinclair D C and West A R 1989 {\it Journal of Applied Physics} {\bf 66} 3850

\bibitem{17} Jonscher A K 1977 {\it Nature} {\bf 267} 673

\bibitem{18} Kao K C 2004  {\it Dielectric Phenomena in Solids} (Elsevier Academic Press) 

\bibitem{19} Costa MM, Pires Jr GFM, Terezo A J, Graca MPF and Sombra ASB 2011 {\it Journal of Applied Physics} {\bf 110} 034107

\bibitem{20} Shamim Md, Singh A and Sharma S 2017 {\it Journal of Advanced Dielectrics} {\bf 7} 1750037

\bibitem{21} Kaushal A, Olhero SM, Singh B, Fagg D P, Bdikin I and Ferreira JMF 2014 {\it Ceramics International} {\bf 40} 10593

\bibitem{22} Singh B K and Kumar B 2010 {\it Crystal Research and Technology} {\bf 45} 1003

\bibitem{23} Nan C-W, Shen Y and Ma Jing 2010 {\it Annual Review of Materials Research} {\bf 40} 131

\bibitem{24} Shetty H D, Patra A and Prasad V 2018 {\it Materials Letters} {\bf 210} 309

\bibitem{25} Shi Z, Mao F, Wang J, Fan R and Wang X 2015 {\it RSC Advances} {\bf 5} 107307

\bibitem{26} Morrison F D, Sinclair D C and West A R 2001 {\it Journal of the American Ceramic Society} {\bf 84} 531

\bibitem{27} Liu J, Duan C, Mei W, Smith R W and Hardy J R 2005 {\it Journal of applied Physics} {\bf 98} 093703

\bibitem{28} Sinclair D C, Adams T B, Morrison F D and West A R 2002 {\it Applied Physics Letterss} {\bf 80} 2153

\bibitem{29} Sivakumar N, Narayanasamy A, Chinnasamy C N and Jeyadevan B 2007 {\it Journal of Physics: Condensed Matter} {\bf 19} 386201

\bibitem{30} Yadav R S, Ku{\v{r}}itka I, Vilcakova J, Havlica J, Masilko J, Kalina L, Tkacz J,  {\v{S}}vec J, Enev V and Hajd{\'{u}}chov{\'{a}} M 2017 {\it Advances in Natural Sciences: Nanoscience and Nanotechnology} {\bf 8} 045002

\bibitem{31} Sharma J, Parashar J, Saxena V K, Bhatnagar D and Sharma K B 2015 {\it Macromolecular Symposia} {\bf 357} 38 

\bibitem{32} Koops, C G 1951 {\it Phys. Rev.} {\bf 83} 121

\bibitem{33} Devan R S, hakras D R, Vichare T G, Joshi A S, Jigajeni S R, Ma Y-R and Chougul B K 2008 {\it Journal of Physics D: Applied Physics} {\bf 41} 105010

\bibitem{34} Ponpandian N, Balaya P and Narayanasamy A 2002 {\it Journal of Physics: Condensed Matter} {\bf 14} 3221

\bibitem{35} Kambale R C, Shaikh P A, Bhosale C H, Rajpure K Y and Kolekar Y D 2009 {\it Smart Materials and Structures} {\bf 18} 115028

\bibitem{36} Murthy V R K and Sobhanadri J 1976 {\it physica status solidi (a)} {\bf 36} K133

\bibitem{37} Kakade S G, Ma Y-R, Devan R S, Kolekar Y D, Ramana C V 2016 {\it The Journal of Physical Chemistry C} {\bf 120} 5682

\bibitem{38} Novotny L and Hecht B 2012 {\it Principles of Nano-optics} (Cambridge University Press)

\bibitem{39} Zhang X, Yan X, He Q, Wei H, Long J, Guo J, Gu H, Yu J, Liu J, Ding D and others 2015 {\it ACS applied materials \& interfaces} {\bf 7} 6125 

\bibitem{40} Shetty H D and Prasad V 2017 {\it Materials Chemistry and Physics} {\bf 196} 153

\bibitem{41} Gao M, Shi Z, Fan R, Qian L,Zhang Z and Guo J 2012 {\it Journal of the American Ceramic Society} {\bf 95} 67

\bibitem{42} Guerra J S and Eiras J A 2007 {\it Journal of Physics: Condensed Matter} {\bf 19} 386217

\bibitem{43} Reddy H, Guler U, Kildishev A V, Boltasseva A and Shalaev V M 2016 {\it Opt. Mater. Express} {\bf 6} 2776

\bibitem{44} Iyengar, Leela and Prasad, B Ram and Qadri, Burhanullah 1973 {\it Current Science} {\bf 42} 534

\bibitem{45} Sundari S T, Chandra S, Tyagi AK 2013 {\it Journal of Applied Physics} {\bf 114} 033515

\bibitem{46} Md A R, Md H, Md S I and Md A R S 2015 {\it APPLIED RESEARCH JOURNAL} {\bf 1} 127

\bibitem{47} Azizian-Kalandaragh Y and Alizadeh-Siakeshi N 2014 {\it JOURNAL OF OPTOELECTRONICS AND ADVANCED MATERIALS} {\bf 16} 345

\bibitem{48} Pendry J B, Holden A J, Stewart W J and Youngs I 1996 {\it Phys. Rev. Lett.} {\bf 76} 4773

\bibitem{49} Chen W, Zhu W, Tan O K and Chen X F 2010 {\it Journal of Applied Physics} {\bf 108} 034101

\bibitem{50} Sivakumar N, A Narayanasamy A, Jeyadevan B, Justin Joseyphus R and Venkateswaran C 2008 {\it Journal of Physics D: Applied Physics} {\bf 41} 245001

\bibitem{51} Shi Z C, Fan R H, Zhang Z D, Gong H Y, Ouyang J, Bai Y J, Zhang X H  and Yin L W  2011 {\it Applied Physics Letters} {\bf 99} 032903

\bibitem{52} Yousefi N, Sun X, Lin X, Shen X, Jia J, Zhang B, Tang B, Chan M and Kim JK 2014 {\it Advanced Materials} {\bf 26} 5480

\bibitem{53} Yu K, Bai Y, Zhou Y, Niu Y and Wang H 2014 {\it Applied Physics Letters} {\bf 104} 082904

\bibitem{54} Yu K, Wang H, Zhou Y, Bai Y and Niu Y 2013 {\it Journal of applied physics} {\bf 113} 034105

\bibitem{55} Pelaiz-Barranco A, Gutierrez-Amador MP, Huanosta A and Valenzuela R 1998 {\it Applied physics letters} {\bf 73} 2039

\bibitem{56} Funke K 1993 {\it Progress in Solid State Chemistry} {\bf 22} 111

\bibitem{57} Ortega N, Kumar A, Bhattacharya P, Majumder S B and Katiyar R S 2008 {\it Phys. Rev. B} {\bf 77} 014111

\bibitem{58} Kumari S, Ortega N, Kumar A, Pavunny S P, Hubbard J W, Rinaldi C, Srinivasan G, Scott J F and Katiyar Ram S 2015 {\it Journal of Applied Physics} {\bf 117} 114102

\bibitem{59} Almond DP, West AR and Grant R\_J 1982 {\it Solid State Communications} {\bf 44} 1277

\bibitem{60} Elliott SR 1977 {\it Philosophical Magazine} {\bf 36} 1291

\bibitem{61}  Austin I G and Mott N F 2004 {\it Advances in Physics} {\bf 18} 41

\bibitem{62} Ghosh A 1990 {\it Phys. Rev. B} {\bf 41} 1479





\end{thebibliography}
\end{document}